\documentclass[twocolumn,amsmath,amssymb,prl,10pt,nofootinbib,superscriptaddress]{revtex4}
\usepackage{ graphicx, float,amsmath, amsmath, amssymb}
\usepackage[export]{adjustbox}

\def\be{\begin{equation}}
\def\ee{\end{equation}}
\def\bea{\begin{eqnarray}}
\def\eea{\end{eqnarray}}
\def\bse{\begin{subequations}}
\def\ese{\end{subequations}}

\usepackage[breaklinks, colorlinks, citecolor=blue]{hyperref}
\usepackage[labelsep=period]{caption}
\usepackage[normalem]{ulem}
\usepackage{mathtools}
\usepackage{siunitx}
\usepackage{soul}
\usepackage{minitoc}

\usepackage{bm}

\DeclarePairedDelimiterXPP\BigOSI[2]%
  {\mathcal{O}}{(}{)}{}%
  {\SI{#1}{#2}}

\begin{document}
\title{Catastrophic fate of Schwarzschild black holes in a thermal bath}

\author{Aurélien Barrau}%
\affiliation{%
Laboratoire de Physique Subatomique et de Cosmologie, Univ. Grenoble-Alpes, CNRS/IN2P3\\
53, avenue des Martyrs, 38026 Grenoble cedex, France
}


\author{Killian Martineau}%
\affiliation{%
Laboratoire de Physique Subatomique et de Cosmologie, Univ. Grenoble-Alpes, CNRS/IN2P3\\
53, avenue des Martyrs, 38026 Grenoble cedex, France
}

\author{Cyril Renevey}%
\affiliation{%
Laboratoire de Physique Subatomique et de Cosmologie, Univ. Grenoble-Alpes, CNRS/IN2P3\\
53, avenue des Martyrs, 38026 Grenoble cedex, France
}




\newcommand{\edit}[1]{{\color{red}{#1}}}

\date{\today}
\begin{abstract} 
Using the Schwarzschild metric as a rudimentary toy model, we pedagogically revisit the curious prediction that the mass of a classical black hole in a constant temperature thermal bath diverges in a finite amount of time. We study in detail how this instability behaves if the temperature of the bath is allowed to vary with time and conclude that whatever the background behavior (but for a zero-measure subspace of the initial conditions), the black hole mass either diverges or vanishes in a finite time if the Hawking radiation is taken into account. The competition between both effects is subtle and not entirely governed by the hierarchy of the relevant temperatures. This instability is also shown to be reached before the background singularity in a contracting universe, which has implications for bouncing models. The results are generalized to spaces with extra dimensions and the main conclusions are shown to remain true. The limitations of the model are reviewed, both from the point of view of the dynamical black hole horizon and from the point of view of the background space expansion. Comparisons with other approaches are suggested and possible developments are underlined.
\end{abstract}
\maketitle

\section{Introduction}

This work does not contain fundamentally new results. It basically aims at pedagogically ``rediscovering" in a wider way a strange behavior of black holes absorbing a continuous flux of energy and at exploring interesting situations relevant for cosmology. Most of our points are established using the Schwarzschild metric in a  Bondi–Hoyle–Lyttleton basic approximation scheme, which can only be considered as a very rough toy model. Interesting features can be guessed while neglecting both the background space expansion and the truly dynamical nature of the black hole horizon but extreme care should be taken when considering the results literally: technical and conceptual issues should be addressed before firm conclusions are drawn. Our aim is therefore not to make any strong claim but to encourage further studies so as to clarify interesting features and to attract the attention of the unfamiliar reader to some strange situations.\\

A classical black hole in a thermal bath obviously grows by absorbing the surrounding radiation. If the temperature of the bath is constant, it might be naively expected that its mass tends to infinity after an infinite amount of time. We show that, without any exotic assumption, the mass actually diverges at {\it finite} time. This has first been noticed in \cite{Novikov2} (translated in \cite{Novikov}). Many works were devoted to the so-called self-similar solution, to the investigation of different equations of state, to the existence of a Friedmann or quasi-Friedmann asymptotic behavior, and to the separate universe issue \cite{Carr:1974nx,Carr:1975qj,bick,carr,Maeda:2002vq,lin,hacyan,Madsen:1988ph,Harada:2004pe,Harada:2006dv,Maeda:2007tk,Kyo:2008qi,Carr:2010wk,Kopp:2010sh,Carr:2014pga}. Although those points are of unquestionable importance, we do not here deal with these subtleties and mostly focus on some strange consequences of a na\"{\i}ve Schwarzschild-based analysis. Our aim is not to derive reliable conclusions -- as a static metric is used beyond its regime of validity -- but to underline some (maybe) surprising situations that could deserve a closer look and seem, to the best of our knowledge, quite ignored by the community.\\

We show that the pathological behavior depends neither on the kind of radiation nor on the initial black hole mass. If the background varies with time, the situation becomes intricate. Depending both on the speed of the space dilation and on the initial conditions, the mass of the black hole can either diverge in a finite amount of time or tend to a finite asymptotic value. The ``natural" expectation ($M\rightarrow \infty$ for $t\rightarrow \infty$) actually happens only for a subset of zero measure in the parameter space (which has already received a great deal of attention in the literature and will not be our focus).\\ 

We compare the critical black hole mass with the Hubble mass and show that different hierarchies have to be considered. When the Hawking evaporation is also taken into account, the mass can vanish for a part of the possible initial states. This shows that, under the strong simplifications performed, in all cases, the black hole is unstable and its mass either diverges or vanishes in a finite amount of time. The relative value between the initial temperature of the bath and the initial temperature of the black hole does {\it not} determine the long run behavior.\\

Importantly, in contracting cosmological solutions, the black hole instability is always reached before the background singularity. Some consequences for bouncing models are drawn, underlying that this catastrophic behavior can happen before the Planck energy is reached. The competition between the absorption and the evaporation is quite involved.\\

We study how extra dimensions would influence the picture and conclude that the diverging behavior remains true although the singularity is reached later when $D>4$.\\ 

Finally, some comparisons with analog black holes are suggested and links with chemistry and statistical physics are pointed out. The limitations of the toy model used in this work are  discussed in detail and possible developments are underlined, focusing on some relevant generalized metrics. We stress that the curious results -- some already well known, others quite new -- shown in this work are more to be understood as a strangeness of the Schwarzschild solution pushed beyond what it was intended for than as realistic physical effects. Still, they can be useful to discover situations deserving deeper investigations.\\

\section{Absorption by a black hole in a thermal bath at constant temperature}

To fix ideas, let us begin by considering a black hole in a constant temperature thermal bath of photons, disregarding the Hawking evaporation. Throughout the article we use Planck units except otherwise stated. The energy density of the equilibrium distribution is given by
\begin{equation}
\rho=\frac{g}{2\pi^2}\int_0^{\infty}\frac{E}{e^{\frac{E-\mu}{T}}-1}E^2dE,
\end{equation}
where $g$ is the number of internal degrees of freedom, $E$ is the energy and $\mu$ is the chemical potential assumed to be negligible in the following. This leads to the standard relation: 
\begin{equation}
\rho=\frac{\pi^2}{15}T^4.
\label{ener}
\end{equation}
The radius of a static\footnote{Once again, we emphasize that this hypothesis is not fulfilled in our calculation, hence the ``toy model" qualification.} nonspinning and uncharged black hole of mass $M$ is $2M$. Photons will fall in the black hole if they approach its center at a distance smaller that $3M$ in Schwarzschild coordinates. This corresponds to an impact parameter at infinity $b=\sqrt{27}M$. Assuming that the wavelength of the background radiation is much smaller than the size of the black hole, that is assuming that the optical limit holds, the effective cross section is simply given by $108M^2$ (the factor $\pi$ being lost by averaging over the isotropic distribution). The mass evolution therefore immediately reads
\begin{equation}
\frac{dM}{dt}=\frac{36}{5}\pi^2T^4M^2.
\label{firstdyn}
\end{equation}
The energy density given by Eq. (\ref{ener}) and entering Eq. (\ref{firstdyn}) is defined at infinity. The equation is trivially integrated in
\begin{equation}
\frac{1}{M_i}-\frac{1}{M}=\frac{36}{5}\pi^2T^4t,
\end{equation}
where $M_i$ is the initial mass of the black hole and $t$ is time elapsed since its formation. Interestingly (and maybe surprisingly), the mass does not diverge in the limit $t\rightarrow \infty$ but at {\it finite} time:
\begin{equation}
t_d=\frac{5}{36\pi^2T^4M_i},
\end{equation}
such that
\begin{equation}
\lim\limits_{t \to t_d}M=\infty.
\end{equation}
This basically means that any black hole in a thermal bath, described at this level of approximation, is unstable. (Alternatively, this also means, with the same limitations, that any steady-state cosmological scenario -- relying on an infinitely old universe and trying to account for a blackbody radiation usually assumed to be at constant temperature -- is basically incompatible with the existence of black holes in the usual sense.)\\

This result remains true if the initial mass of the black hole is such that the wavelength of the surrounding radiation is much larger than the Schwarzschild radius. In this case, the scattering cross section is given by \cite{MacGibbon:1990zk}:
\begin{equation}
\sigma=\frac{64\pi M^4E^2}{3}.
\end{equation}
This new behavior is due to the fact that the incoming wave cannot be approximated to a point particle anymore. The additional $E^2$ factor shows that the greybody factor is, as expected, suppressed in the limit $ME\rightarrow 0$. In this regime, and assuming $E\sim T$, the evolution equation becomes
\begin{equation}
\frac{dM}{dt}=\frac{256}{45}\pi^2T^6M^4.
\label{cross_low_energy}
\end{equation}
Calling $\lambda$ the mean wavelength of the thermal radiation, we define the equilibrium time (corresponding to the transition between a black hole smaller than the typical photons to a black hole larger than the typical photons) $t_e$ such that $M(t_e)\sim \lambda \sim 1/T$. This defines the change of regime between the optical cross section $\propto M^2$ and the low-energy cross section $\propto M^4E^2$. Neglecting the oscillations (which would not change the order of magnitude), $t_e$ can be estimated to be given by integrating Eq.(\ref{cross_low_energy}):
\begin{equation}
t_e=\frac{15}{256\pi^2T^6}\left(M_i^{-3}-T^3\right),
\end{equation}
with $M_i<1/T$ by hypothesis.


After the time $t_e$, the black holes are in the usual regime and the remaining time before divergence is
\begin{equation}
\Delta t=\frac{5}{36\pi^2T^3}.
\end{equation}

The full time between formation and divergence is therefore in this case
\begin{eqnarray}
t_d&=&\frac{5}{36\pi^2T^3}+\frac{15}{256\pi^2T^6}\left(M_i^{-3}-T^3\right)\\
&=&\frac{186}{2304\pi^2T^3}+\frac{15}{256\pi^2T^6M_i^3}
\end{eqnarray}
which is again finite for any value of $M_i$ and $T$. The dependence upon the initial mass is, as expected, stronger than before as the absorption at the beginning is highly suppressed. This phenomenon, however, does not prevent the mass divergence at finite time. \\

If the thermal bath is made of relativistic fermions instead of photons, the result is mostly the same in the case of a black hole larger that the mean wavelength of the radiation. The energy density of the bath is simply modified by a factor of $7g/16$ and

\begin{equation}
t_d=\frac{20}{63\pi^2gT^4M_i}.
\end{equation}

The ``small mass regime", however, strongly differs from the case of photons. The scattering cross section for fermions in the $E\rightarrow 0$ limit is \cite{MacGibbon:1990zk}
\begin{equation}
\sigma=2\pi M^2.
\end{equation}
The same reasoning as previously leads to
\begin{equation}
t_e=\frac{30}{7\pi^2g T^4}\left( M_i^{-1}-T \right).
\end{equation}
And the full divergence time reads
\begin{eqnarray}
t_d&=&\frac{20}{63\pi^2gT^3}+\frac{30}{7\pi^2g T^4}\left( M_i^{-1}-T \right)\\
&=&\frac{30}{7\pi^2gT^4M_i}-\frac{256}{63\pi^2gT^3}.
\end{eqnarray}
This shows that, in all cases, the mass of a classical black hole in a constant temperature bath tends to infinity in a finite amount of time.

This unusual behavior is, of course, entirely rooted in the specific mass-radius relation of black holes. A ball of standard matter with a cross section proportional to $M^{2/3}$ would grow gently as $M\propto t^3$ and would never experience any singularity.

\section{Absorption in a thermal bath at decreasing temperature}

Although the mass divergence in finite time might come, at first sight, as a physical surprise, it is a mathematically obvious consequence of having $dM/dt\propto M^{\delta}$ with $\delta>1$. It is now worth considering the case of a classical black hole immersed in a thermal bath with decreasing temperature. Let us assume that 
\begin{equation}
T=T_0\left( \frac{t}{t_0} \right)^{\alpha}.
\end{equation}
The constants $T_0$ and $t_0$ could be absorbed in a single parameter but keeping both of them helps the physical intuition. The exponent $\alpha$ is negative (otherwise the divergence is just trivially amplified). The evolution equation (focusing on the case of an initial mass larger than the radiation mean wavelength) is
\begin{equation}
\frac{dM}{dt}=kT^4M^2=kT_0^4\left( \frac{t}{t_0} \right)^{4\alpha} M^2,
\end{equation}
with $k=36\pi^2/5$ for photons. This leads to
\begin{equation}
\frac{1}{M}=\frac{1}{M_i}-\frac{kT_0^4}{(4\alpha+1)t_0^{4\alpha}} \left( t^{4\alpha +1} - t_i^{4\alpha +1} \right),
\end{equation}
where $t_i$ is the formation time of the black hole and $M_i$ its corresponding mass. Calling $\beta=4\alpha+1$, the mass diverges $(1/M=0)$ at time
\begin{equation}
t_d=e^{\frac{1}{\beta}ln \left( \frac{\beta t_0^{\beta-1}}{kM_iT_0^4} +t_i^{\beta} \right)}.
\label{div}
\end{equation}
This will actually happen if the argument of the logarithm is positive. If $\beta >0$, this is always true. Otherwise stated, if the cooling of the universe in slow enough ($T\propto t^{\alpha}$ with $\alpha >-1/4$), the mass divergence at finite time happens whatever the initial conditions. On the other hand, if $\beta <0$, the divergence requires $M_i>M_c$ with
\begin{eqnarray}
M_c&=&-\frac{\beta t_0^{\beta -1}t_i^{-\beta}}{kT_0^4}\\
&=&\left( \frac{8}{3(1+w)}-1 \right) \frac{t_0^{-\frac{8}{3(1+w)}}}{kT_0^4}t_i^{\left( \frac{8}{3(1+w)-1} \right)}.
\label{betaneg}
\end{eqnarray}

In a cosmological setting, $T\propto a^{-1}$ -- with $a$ the scale factor -- that is, $T\propto t^{-\frac{2}{3(1+w)}}$ with $w=p/\rho$ the equation of state parameter for the dominant fluid. The condition $\beta >0$, or equivalently $\alpha>-1/4$, translates into $w>5/3$. This value, greater than one, corresponds to ``superstiff" matter. Although quite exotic, this behavior can be encountered in realistic models such as Horava-Lifshitz gravity (see \cite{KalyanaRama:2009px}). In this case, even if the temperature is decreasing, the black hole mass diverges whatever its initial value. If $\alpha<-1/4$, the divergence is associated with the condition given by Eq. (\ref{betaneg}). The important point is that there always exists an initial mass beyond which, the divergence does occur. In this sense, part of the parameter space is unavoidably unstable. \\

It is interesting to compare $M_c$ with the Hubble mass $M_H$. At time $t_i$, the latter is of order $M_H\sim t_i$. The condition $M_c<M_H$ is always fulfilled when $\beta>0$, which means that the considered black hole can form without any causality issue. On the other hand, if $\beta<0$, the condition reads
\begin{equation}
t_i^{-\beta-1}<\frac{kT_0^4}{-\beta t_0^{\beta -1}}.
\end{equation}
Let us define the critical crossing time
\begin{equation}
t_{cH}=\left( \frac{kT_0^4}{-\beta t_0^{\beta-1}} \right)^{\frac{-1}{\beta+1}}.
\end{equation}
If $-\beta-1>0$ (that is, $\alpha<1/2$ or $w<1/3$), the condition translates into $t_i<t_{cH}$ whereas if $-\beta-1<0$ (that is, $\alpha>1/2$ or $w>1/3$), the condition translates into $t_i>t_{cH}$. For content with an equation of state softer than radiation, unstable (with diverging mass) black holes can be causally formed early in the history of the Universe, while for a background equation of state stiffer than radiation, unstable black holes can form late in the cosmological history. If $w=1/3$ exactly, the condition simply reads $-\beta t_0^{\beta -1}<kT_0^4$. In all cases a part of the parameter space leads to diverging black holes. The detailed investigation of hierarchy between the horizon of the black hole and the cosmological horizon (particle horizon before the inflation, Hubble horizon after the inflation) has been extensively studied, {\it e.g.} in \cite{Carr:1974nx,carr,Harada:2004pe}, and we will not repeat the analysis here.\\

\begin{figure}
    \centering
    \includegraphics[width=1\linewidth]{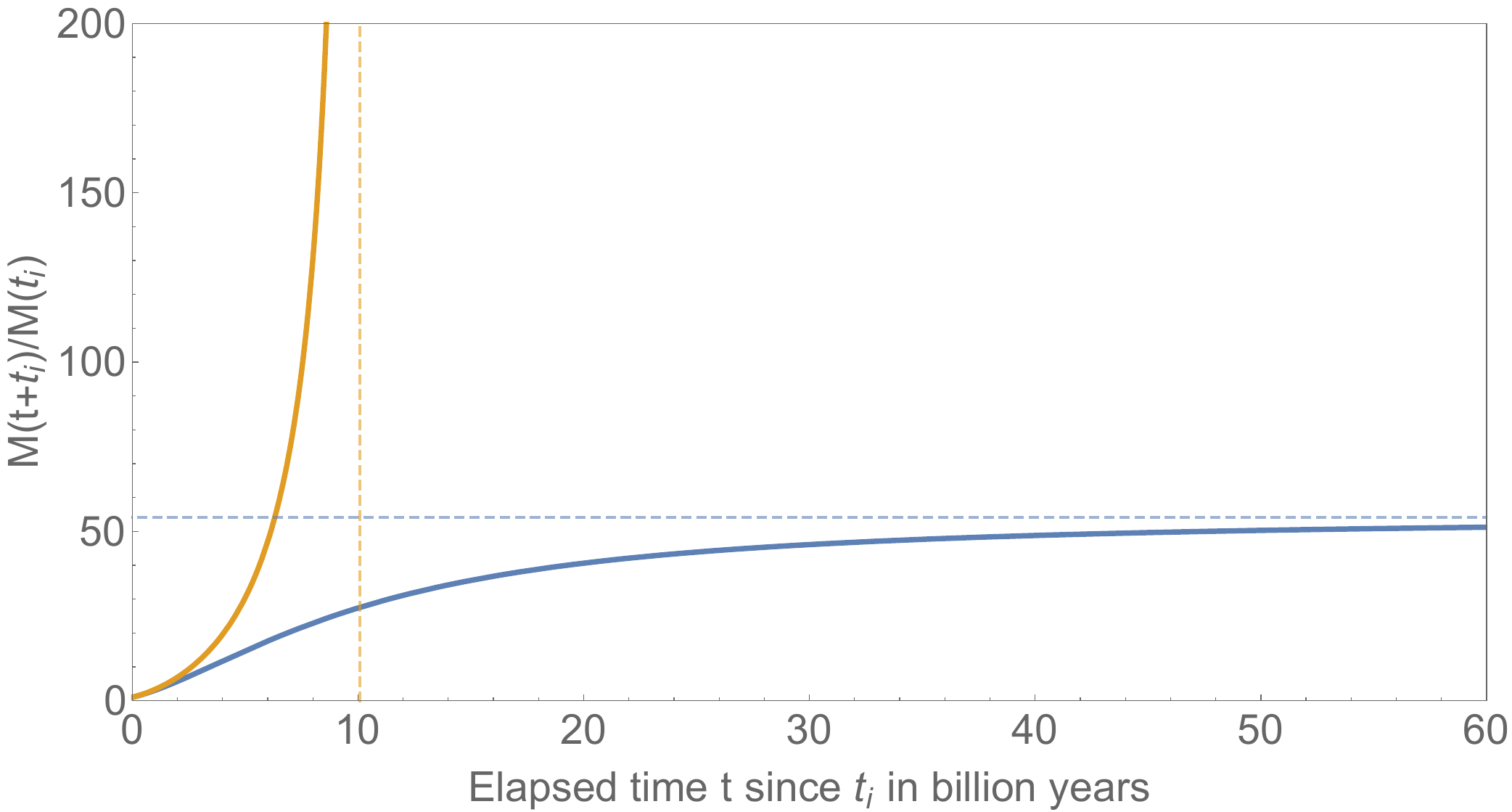}
    \caption{Evolution of the mass of a black hole in a perfect fluid with $w=0$. The blue curve corresponds to a black hole with an initial mass slightly lower than the critical mass $M_c$ whereas the yellow curve corresponds to a mass slightly higher. Dashed lines represent asymptotic behaviors. No ``in-between" dynamics is possible.}
    \label{fig:no_evap}
\end{figure}

Interestingly, the naively expected behavior ($M\rightarrow \infty$ for $t\rightarrow \infty$), that is also the one which has generated some interest from the point of view of general relativity, happens only for a zero measure parameter space. In the general case where the instability is avoided, the mass tends to a finite asymptotic value $M_{\infty}$ in the remote future:
\begin{equation}
M_{\infty}=\left( \frac{1}{M_i}+\frac{kT_0^4t_i^{\beta}}{\beta t_0^{\beta -1}}  \right)^{-1}.
\end{equation}
This is illustrated in Fig. \ref{fig:no_evap}. Depending on the initial mass and on the speed at which the temperature of the thermal bath decreases, the evolution of the black hole mass corresponds to one of the two cases displayed. 

\section{Switching on the evaporation}

Obviously, the full picture requires one to also take into account the Hawking evaporation \cite{Hawking:1975vcx} which is mandatory to have a consistent thermodynamical understanding (although the truly dynamical horizon is still ignored). The evolution equation now reads
\begin{equation}
\frac{dM}{dt}=kT^4M^2-\gamma M^{-2}.
\end{equation}
In principle, the $\gamma$ parameter depends on the mass $M$ as the number of degrees of freedom available increases with the black hole temperature $T_{BH}=1/(8\pi M)$. New channels are opened each time the temperature becomes higher than the rest mass of a given particle species. For this study it is clearly sufficient to assume $\gamma$ to be constant \cite{Halzen:1991uw}. Its numerical value can be straightforwardly calculated by integrating the Hawking spectrum (multiplied by the energy $Q$ of the emitted particle):
\begin{equation}
\frac{dM_{evap}}{dt}=-\int\frac{\Gamma}{2\pi}\left( e^{\frac{Q}{T_{BH}}} -(-1)^{2s}) \right)^{-1}QdQ,
\end{equation}
where $s$ is the spin of the particle and $\Gamma=Q^2\sigma/\pi$ is the greybody factor. \\

The Hawking evaporation has an immediate consequence. If the black hole happens to be in an initial state where the evaporation dominates over the absorption, it automatically remains in this regime: the mass decreases and the $M^{-2}$ term becomes more and more important with respect to the absorption one. The situation considered at the end of the first section is therefore purely academic: in practice, if the size of the black hole is smaller than the wavelength of the surrounding radiation, the black hole behavior is dominated by the evaporation. Its mass vanishes in a time

\begin{equation}
t_{evap}=\frac{M_i^3}{3\gamma}.
\end{equation}
This means that, in a thermal bath, a black hole is anyway unstable: its mass either diverges or reaches zero\footnote{There are countless arguments and models in quantum or extended gravity to avoid the naked singularity - see references in \cite{Barrau:2019cuo} - but this is not the point we wish to make here.} in a finite amount of time. The Hawking time scales as $M_i^3$ and the divergence time scales as $M_i^{-1}$.\\

\begin{figure}
    \centering
    \includegraphics[width=1\linewidth]{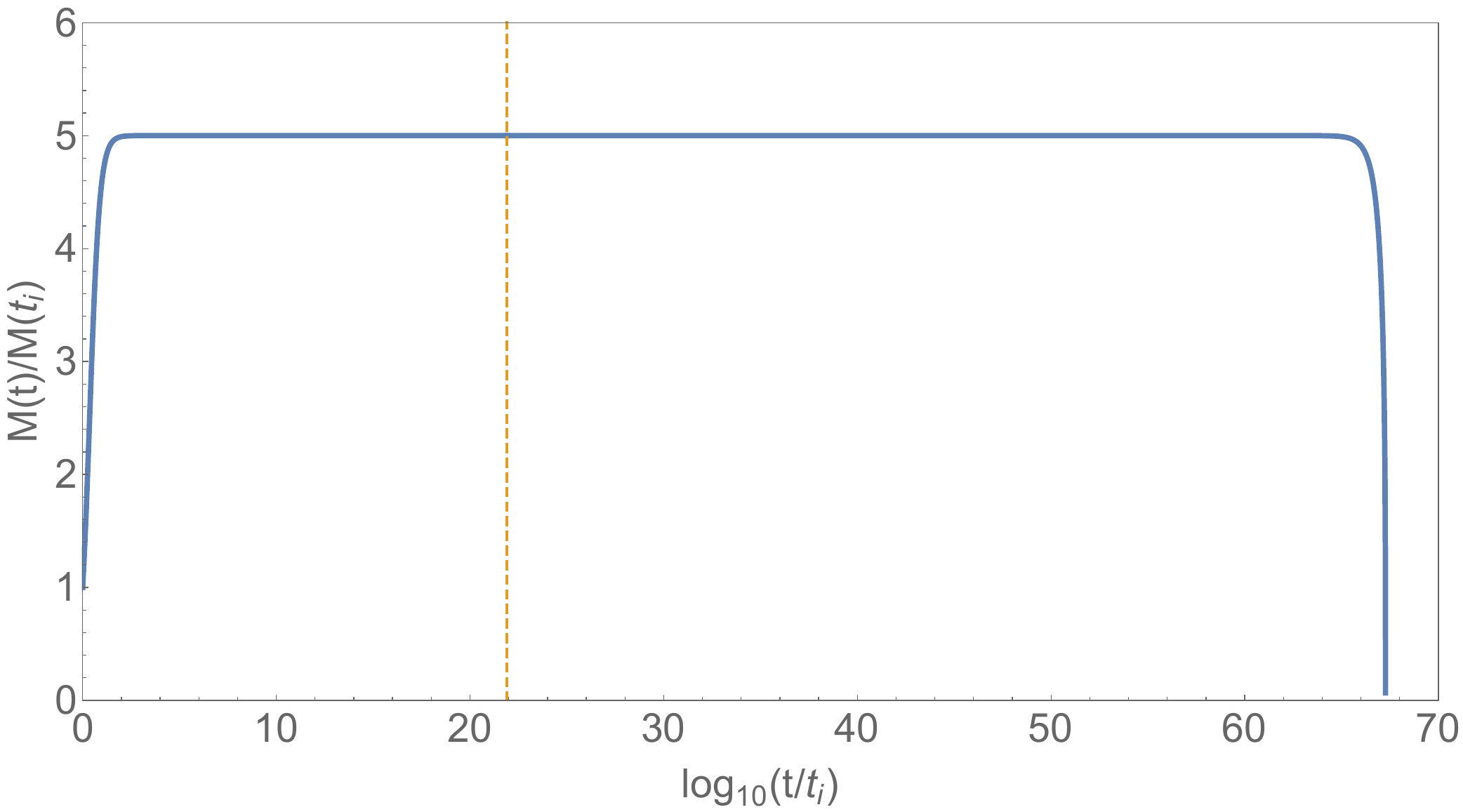}
    \caption{Evolution of the mass of a black hole in a perfect fluid with $w=0$ taking into account Hawking radiation. The blue curve corresponds to a black hole with an initial mass slightly lower than the critical mass $M_c$ whereas the yellow dashed line corresponds to the time when evaporation starts to dominate.}
    \label{fig:with_evap}
\end{figure}

The reciprocal of the previous statement is, however, not true in an expanding universe. If the absorption initially dominates, this does not mean that it will remain so forever. Actually, it {\it cannot} remain true in most cases.
If the cooling of the Universe is ``usual" ({\it e.g.} associated with an equation of state $w=1/3$ or $w=0$) and the initial black hole mass is such that the divergence is avoided ($M_i<M_c$), the evaporation will inevitably dominate at some point. In the absorption-dominated regime, the mass of the black hole increases and its temperature decreases. However, as previously shown, the mass necessarily tends to a finite value in the remote future. This means that, as far as absorption is concerned, $dM/dt \rightarrow 0$. On the other hand, the mass variation due to the evaporation has a constant asymptotic value $dM/dt=-\gamma M_{\infty}^{-2}$. After some time $t_*$ this latter term will dominate. This basically means that the naive idea according to which, in a dynamical thermal bath, a black hole either absorbs radiation or evaporates depending only on the respective (initial) values of the bath and of the black hole temperatures is wrong. If the initial temperature of the black hole is smaller than the initial temperature of the background, the black hole temperature will first decrease (it will grow by absorption), in accordance with the usual view. However (if the dynamics is not the diverging one), after a finite time $t_*$, the temperature will start to increase again (the black hole will shrink by evaporation). Once the evaporation dominates, the behavior will not reverse until the disappearance of the black hole. The minimum temperature reached by the black hole is 
\begin{equation}
T_{min}=\frac{1}{8\pi}\left( \frac{1}{M_i}+\frac{kT_0^4t_i^{\beta}}{\beta t_0^{\beta -1}}  \right).
\end{equation}
The transition time $t_*$ is such that
\begin{eqnarray}
\sqrt{\frac{\gamma}{k}}\frac{t_0^{2\alpha}}{T_0^2 t_*^{2\alpha}} \left( \frac{1}{M_i}-\frac{kT_0^4}{(4\alpha+1)t_0^{4\alpha}} \left( t_*^{4\alpha +1} - t_i^{4\alpha +1} \right) \right)\\
= \left( M_i^3 -3\gamma (t_*-t_i) \right)^{\frac{1}{3}}.
\end{eqnarray}

If the dynamics of the background spacetime is, however, such that the black hole mass should diverge due to absorption, the situation is quite subtle. For most of the parameter space, the black hole just grows and the evaporation does not play any role. But it can be shown that for highly tuned initial conditions the second time derivative of the mass can vanish and the evaporation can overcome the growth. This corresponds to a very particular case worth being mathematically pointed out but most probably without any phenomenological consequences.\\

\begin{figure}
    \centering
    \includegraphics[width=1\linewidth]{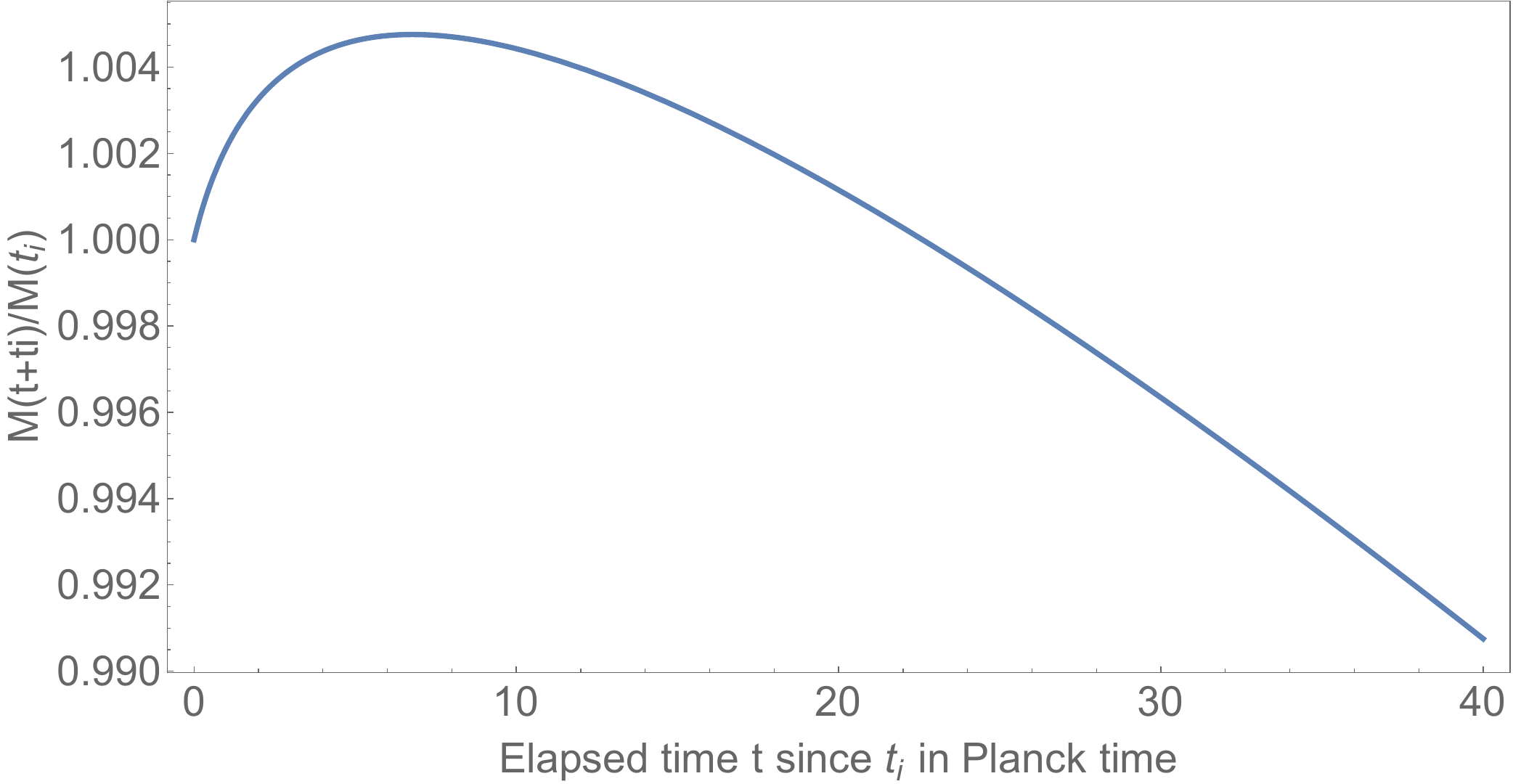}
    \caption{Very special case where the absorption initially dominates in an expanding space -- in a regime such that the mass should diverge without evaporation -- but where the Hawking effects still finally overcome the evolution.}
    \label{fig:inv}
\end{figure}

This shows why black holes in a thermal bath are {\it unstable}. The mass either diverges or vanishes in a finite amount of time. Figure \ref{fig:with_evap} illustrates the typical behavior in the case where the absorption initially dominates without being diverging: the mass first increases quite fast, then remains close to its asymptotic value for most of the evolution, and then decreases until it completely vanishes. The very highly tuned case where the mass would diverge without evaporation but where the dynamics is finally overcome by the Hawking effect is exhibited in Fig \ref{fig:inv}.\\

Let us summarize:
\begin{itemize}
\item If the initial black hole temperature is higher than the background temperature, the black hole simply evaporates and vanishes.
\item If the initial black hole temperature is smaller than the background temperature and $\alpha>-1/4$ or ($\alpha<-1/4$ and $M>M_c$), the black hole mass generically diverges (however, there exists a tiny set of parameters -- corresponding to initial conditions highly tuned close to the critical values -- where the evaporation finally dominates),
\item If the initial black hole temperature is smaller than the background temperature and ($\alpha<-1/4$ and $M<M_c$), the black hole mass vanishes after having reached a ``plateau" where it stayed for the vast majority of its lifetime.
\end{itemize}
The important feature is that the infinite or vanishing mass is always reached in a finite amount of time. Let us mention, although this is not the point of this study, that even if some kind of remnant or relic is formed at the end of the evaporation, they are anyway usually expected to disappear at some point \cite{Giddings:1992hh}.

\section{The de Sitter case}

When the scale factor expands exponentially, the temperature evolution can be written as 
\begin{equation}
T=T_0e^{-\kappa t}.
\end{equation}
The mass evolution (when the black hole initial mass is such that the absorption dominates over evaporation) then reads
\begin{equation}
\frac{1}{M}=\frac{1}{M_i}+\frac{kT_0^4}{4\kappa}\left( e^{-4\kappa t} - e^{-4\kappa t_i} \right).
\end{equation}
As expected, no divergence occurs in this case and the mass always reaches its asymptotic value,
\begin{equation}
M_{\infty}=\left( \frac{1}{M_i}-\frac{kT_0^4}{\kappa}e^{-4\kappa t_i} \right)^{-1}.
\end{equation}
Once again, the evaporation unavoidably dominates at some time $t_*$ such that
\begin{eqnarray}
\sqrt{\frac{\gamma}{k}}kT_0^{-2}e^{2\kappa t_*}\left( \frac{1}{M_i}+\frac{kT_0^4}{4\kappa} (e^{-4\kappa t_*}-e^{-4\kappa t_i}) \right)=\\
 \left( M_i^3 -3\alpha (t_*-t_i) \right)^{\frac{1}{3}}.
\end{eqnarray}
The de Sitter horizon is endowed with a temperature 
\begin{equation}
T_{dS}=\frac{1}{2 \pi}\sqrt{\frac{\Lambda}{3}},
\end{equation}
where $\Lambda$ is the cosmological constant. This, however, does not change the picture in any noticeable way as the black hole temperature always remains higher than the de Sitter temperature. \\

\section{The contracting Universe catastrophe}

Quite a lot of theories beyond general relativity predict a cosmological bounce instead of the Big Bang singularity (see, {\it e.g.}, \cite{Battefeld:2014uga,Lilley:2015ksa,Brandenberger:2016vhg} for reviews). This is even possible in general relativity without exotic matter \cite{Barrau:2020nek,Renevey:2020zdj}. In such models, the Universe was contracting before entering the current expanding branch. If space was, before the bounce, filled with black holes and radiation\footnote{There are {\it no} motivations for this assumption often made in the framework of bouncing models. This is only justified by the desire of studying a symmetric situation that might be the less unjustified guess. It remains hazardous from the causality point of view.}, the catastrophic growth of black hole masses that we have established in a constant-temperature bath will even be worsened.\\

\begin{figure}
    \centering
    \includegraphics[width=1\linewidth]{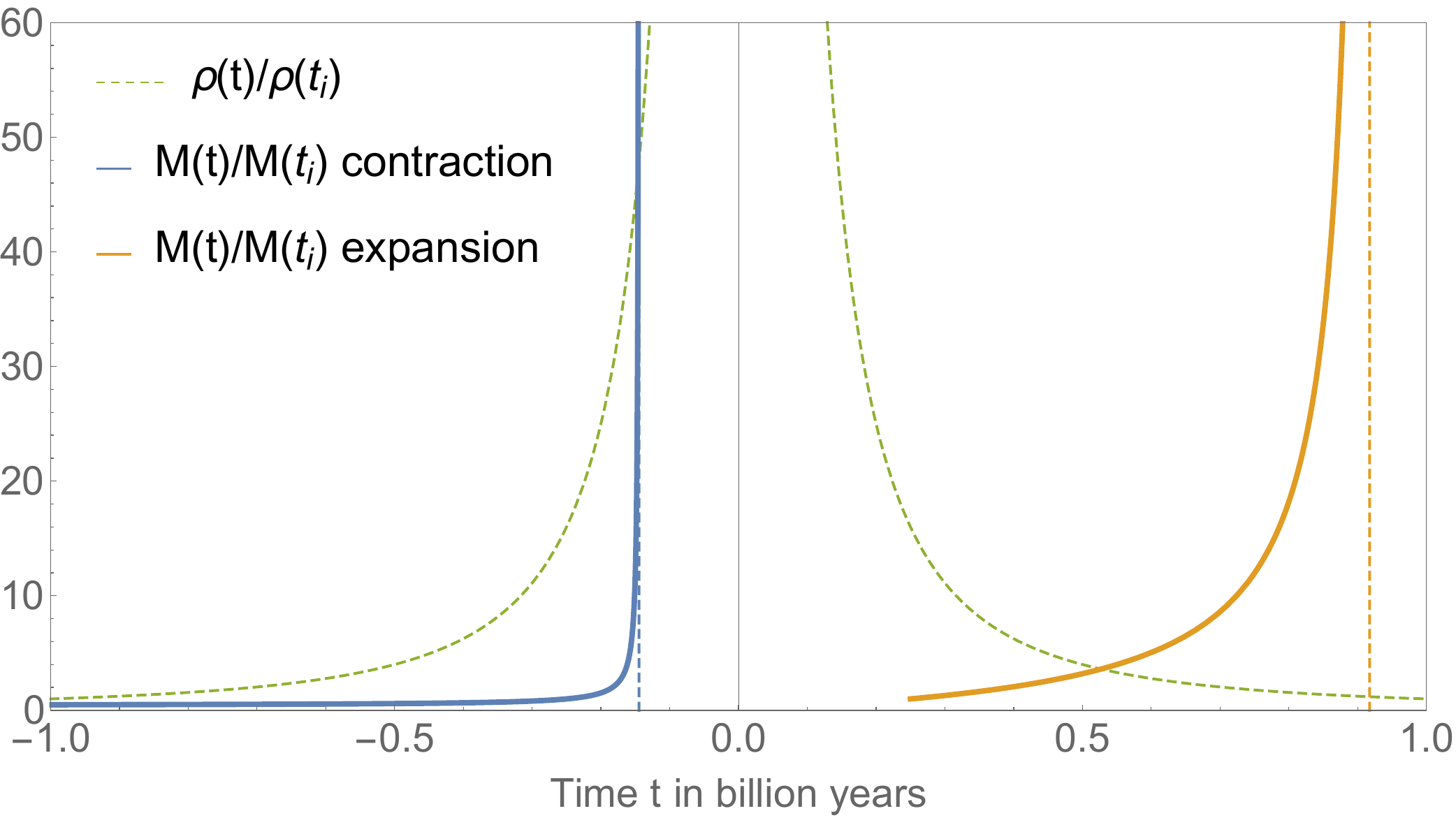}
    \caption{The blue curve in the left panel and the yellow curve in the right panel represent the time evolution of black holes with the same initial masses in a radiation-dominated universe which is, respectively, contracting (left) and expanding (right). The dashed green curves correspond to the evolution of the background density.}
    \label{fig:contr}
\end{figure}

It makes sense to evaluate the time taken by a black hole to reach its absorption singularity. Contracting spaces are known to exhibit some paradoxes (see \cite{Barrau:2017ukm,Barrau:2014kza}) and it is interesting to compare how long it takes for the black hole to become unstable when compared to the time required for the background radiation to reach the Planck density (triggering quantum gravity effects). If the contracting branch is filled with relativistic matter, the dynamics of the black hole reads
\begin{equation}
\frac{1}{M_i}-\frac{1}{M}=kT_0^4t_0^2 \left( \frac{1}{t_i}-\frac{1}{t}  \right),
\label{contraction}
\end{equation}
where $t$ is now negative and the conventions are the same as previously. (Contrary to what is sometimes believed, if a cosmological variable scales as $t^q$ in an expanding universe, it will {\it not} behave as $t^{-q}$ in a contracting one !). Although this expression is formally the same than in the expanding universe, it does induce, due to the fact that $t$ is now negative, a far {\it faster} divergence, all the other parameters being the same, as it can be seen in Fig. \ref{fig:contr}. The background energy density varies as $\rho\propto (-t)^{-2}$, and Fig. \ref{fig:contr} shows that the black hole mass divergence can (depending on the initial conditions) be reached before the energy density becomes Planckian. This  means that the singularity resolution provided by the bounce in quantum gravity models (see, {\it e.g.}, \cite{lqc9}) might not solve this specific problem which should be seriously considered. To be more illustrative, let us once again consider a prebounce universe similar to our expanding one, {\it i.e.} such that the temperature was $T\sim 3\times 10^3$ K at time $t\sim -3\times 10^5$ years before the bounce. Then a stellar mass black hole with an initial mass $M\sim 10$ M$_\odot$ would diverge at time $t_d\sim - 10^{-5}$s when the temperature was $T(t_d)\sim 10^{12}$ K, well before the Planck era.\\

This also leads an important remark. Let us consider the contracting solution to the Friedmann equation:
\begin{equation}
H=-\sqrt{\frac{8\pi G \rho}{3}},
\end{equation}
where $H$ is the Hubble parameter. This describes a universe which reaches a singularity at $t=0$. It is, however, clear from Eq. (\ref{contraction}) that the point $1/M=0$ will inevitably be reached before $t=0$, whatever the initial conditions. The black hole instability precedes the big crunch singularity. As long as there is a black hole in a contracting universe filled with radiation, this phenomenon will take place before the breakdown of the smooth background evolution.\\

In a contracting space, if the absorption initially dominates, the evaporation obviously never plays any role and the mass diverges. What, however, happens if the evaporation is dominant at the formation time? One might na\"{\i}vely expect that a severe competition takes place between absorption (which increases at a given mass due to the increase of the temperature) and evaporation (which increases due to the decrease of the mass). This, however, is not the case in general. One can straightforwardly show that the absorption overcomes the evaporation if the background temperature satisfies
\begin{equation}
T_{back}>\left( \frac{\gamma}{k} \right)^{\frac{1}{4}} \frac{1}{M}.
\end{equation}
The background temperature, however, diverges at $t=0$. This means that -- once again except for a tiny and mostly irrelevant part of the parameter space corresponding to fine-tuned initial conditions -- the Hawking evaporation remains dominant if the Hawking time is smaller than the time available before the crunch (or the bounce). 

\section{Extra dimensions}

Gravity with more than three spatial dimensions is far richer. This is the case for mathematical reasons: the rotation group $SO(D)$ has Cartan subgroup $U(1)^N$ with $N=E\left[D/2\right]$. This is also the case for physical reasons: the radial falloff of the Newtonian potential scales as $r^{D-2}$ whereas the centrifugal barrier does not depend on $D$.
Let us consider a spacetime with $D$ spatial dimensions and define

\begin{equation}
\mu=\frac{16\pi M}{(D-1)\Omega_{D-1}},
\end{equation}
where
$\Omega_{d-1}=2\pi^{D/2}/\Gamma\left(\frac{D}{2}\right)$ is the
area of a unit $(D-1)$-sphere. The generalized Schwarzschild metric reads \cite{Emparan:2008eg,Kanti:2004nr}:

\begin{equation}
ds^2=-\left(1-\frac{\mu}{r^{D-2}}\right)dt^2+\frac{dr^2}{1-
\frac{\mu}{r^{D-2}}}+r^2 d\Omega_{D-1}^2.
\end{equation}
It is straightforward to show that, in this framework, the time evolution of the black hole mass becomes (in a constant temperature background):
\begin{equation}
\frac{dM}{dt}\propto M^{\frac{D-1}{D-2}}.
\end{equation}
This diverges at finite time if $(D-1)/(D-2)>1$, which is always true. However, the higher the number of extra dimensions, the less drastic the divergence becomes. Care should be taken when using Planck units with $D>3$. The full expression of the radius of the horizon is
\begin{equation}
R_D=\frac{1}{\sqrt{\pi}M_*}\left( \frac{M}{M_*} \right)^{\frac{1}{D-2}} \left( \frac{8\Gamma(\frac{D}{2})}{D-1} \right)^{\frac{1}{D-2}}.
\end{equation}
In this expression $M_*$ is the fundamental Planck scale and {\it not} the usual (3+1)-dimensional one. The Hawking temperature of the black hole reads
\begin{equation}
T_{BH_D}=\frac{D-2}{4\pi R_D}.
\end{equation}
The behavior is qualitatively the same as in the four-dimensional case. Whatever the number of extra dimensions the black hole mass diverges at finite time in a static thermal bath. If, however, the temperature of the bath decreases with time and the black hole is not too large at the initial time, the black hole mass will also tend to an asymptotic value. The dynamics will then, at some point, be dominated by the evaporation and the mass will vanish. 

\section{Analogies}

The simple approach we have chosen here is obviously oversimplified. The dynamics of Eq. (\ref{firstdyn}) reads, for the radius of the black hole,
\begin{equation}
\frac{dR}{dt}=\frac{72}{5}\pi^2T^4R^2.
\end{equation}
This means that $dR/dt \rightarrow \infty$ as the critical time is approached. This is also true in the other backgrounds considered as far as the instability occurs. Is this physical? We will come back to this question at a more fundamental level in the next sections and we remain here at the level of the simple consistency of the toy model. Obviously, if a particle of total energy $\epsilon$ falls into a large black hole of radius $R$, it does not make sense to assume that the horizon jumps instantaneously to a perfect sphere whose radius is simply increased by $2\epsilon$. The intricate relaxation procedure through quasinormal modes should, in principle, be considered to accurately describe the time evolution of the shape of the horizon. For obvious causality reasons the horizon cannot globally vary in shape ``faster than light" as a consequence of a local deformation. The situation considered in this work is, however, different. One deals with a nearly continuous background of ingoing energy.\\

The very notion of a event horizon is, in nature, global and teleological (see, {\it e.g.} \cite{Bhattacharya:2017dgr}). It is {\it the boundary of future null infinity}. This, however, does not prevent the change of the horizon shape, at $t+dt$, due to an incoming flux of energy on a black hole preexisting at time $t$, to be {\it locally} determined. Associated conceptual and technical considerations were studied, {\it e.g.}, in \cite{Kopteva:2018ply,Zhao:2018ani}, following \cite{Israel:1966rt}. The growth of the horizon is locally driven by the global past history {\it and} the recent local events. There is, then, no reason to discard a ``faster than light" expansion of the horizon. This is no stranger than a wave crest moving arbitrarily fast: no matter or information is transported. The region from which (in the black hole case) light rays can escape to infinity might, without violating causality, shrink faster than light. This eventually happens when a black hole is fed by a continuous and homogeneous flux of energy. 

Let us consider for example analog black holes (see, {\it e.g.}, \cite{Ge:2010wx}). The position of the horizon is determined both by nonlinear effects and by the environment whereas the speed of sound is an entirely linear property. The speed of the first can therefore be higher than the speed of the second. Concretely, this would clearly be possible with a two-component Bose-Einstein condensate. The sound speed for the two species, respectively $c_1$ and $c_2$, are {\it a priori} different. If they do interact, the position of the horizon for the first will depend on the density of the second (and the other way around). As there are nonlinear solutions -- like grey solitons -- moving close to sound speed, there can be, from the viewpoint of the first species, a horizon moving at a speed close to $c_2$, which can be greater than $c_1$.\\

With the obvious limitations of the metric used, the growth of the horizon at an arbitrary high speed is not inconsistent in the considered framework. This, by the way, also happens when considering the Hawking evaporation where $dR/dt\rightarrow \infty$ when $M\rightarrow 0$. In this case, one could argue that the semiclassical formula should be modified by quantum gravity effects in this regime but the divergence of the speed is not the reason why a better description should be searched for. The phenomenon underlined in this work has no reason not to be possibly real. Although a kind of ``bubble nucleation" (at the speed of light) similar to a tunneling effect from a false vacuum to the true vacuum (see, {\it e.g.}, \cite{Chen:2020lrl}) would also make sense, the considered process can be faster and even more catastrophic.\\

Finally, it is important to notice that the phenomenon exhibited in this article is not unique in physics. One could first think about the disorder correlator. Functional renormalization shows that its curvature explodes at a finite scale $\ell_c$ (developing a cusp at the origin). This can be interpreted through shocks and avalanches \cite{Wiese:2021qpk}. 

Closer to our situation is probably the case of cubic autocatalysis (that is, of order $n=2$). In such chemical processes, one of the reaction products is also a catalyst for the same reaction. This reads $2A+B\rightarrow 3A$. If the rate of the reverse reaction is vanishing (the rate of the forward reaction being $r$) and if the concentration $x_B$ of the species $B$ is kept constant, the number of $A$ particles evolves (in the large $N$ limit) as $\partial N/ \partial t= r x_b N^2$. This is the same formal equation as the one we have considered in this work for a constant temperature bath. In a way, the black hole corresponds to species $A$.

\section{The black hole as an underdensity}

Among others, a curious aspect of the situation considered in this article is the following. When it grows by absorption of the surrounding radiation, a black hole might end being {\it less} dense that the medium in which it is embedded. Obviously, the Schwarzschild solution, which is a vacuum solution to Einstein's equation, is not appropriate to handle this situation. The very meaning of the ADM mass defining the families of solutions is ill-defined. It is nevertheless worth considering a simple thought experiment to understand whether it is possible to make sense out of this. An interesting case is the one of thick shells collapsing toward a preexisting black hole \cite{Liu:2009ts}. The exact solution was first derived using comoving coordinates, 
\begin{equation}
ds^2 =d\tau^2-e^{\bar{\omega}(R,\tau)}dR^2-e^{\omega (R,\tau)}(d\theta^2+sin^2\theta d\phi^2),
\end{equation}
where there is only one nonvanishing component in the stress-energy tensor. Using the machinery developed by Oppenheimer \cite{Oppenheimer:1939ue} and appropriate matching conditions, the result can be transformed to usual coordinates and is easy to interpret. If several shells collapse on the black hole, one can show that all the incoming matter but the outermost layer of the last one do cross the horizon \cite{Liu:2009ts} (exact solutions for null fluid collapse were obtained in \cite{Husain:1995bf}). This does not rely on any assumption about the respective densities of the black hole and of the shells. The latter can be  denser than the former (the mean density of a black hole is anyway not a physically relevant quantity). It should be kept in mind that we do not consider here accretion of dust, which would be severely impacted, but absorption of radiation whose penetration in the black hole is (mostly) not driven by gravitational effects.\\

Obviously, a black hole does not form in a static homogeneous space without a triggered gravitational instability. Otherwise, one would be led to the absurd conclusion that black holes spontaneously appear in any large enough static space as a region of size $R$ falls inside its own gravitational radius as long as $R>\sqrt{3/(8\pi\rho)}$. This is clearly wrong as this analysis would only be correct for the vacuum case. However, once the black hole is formed there is no reason to discard the possibility that its mean density becomes smaller than the average one of the surrounding medium. Although unusual, this situation is not impossible.

\section{Full dynamics}

Two crucial dynamical aspects, beyond the scope of this article, should be accounted for before any reliable conclusion can be drawn: the expansion (or contraction) of the Universe beyond its consequence on the radiation density and the evolution of the black hole horizon itself. \\

Many works are devoted to both aspects but it remains hard to get a clear and noncontroversial picture. Black holes in a radiation-dominated universe were, in particular, studied in detail in \cite{Babichev:2018ubo}.  The metric obtained is
\begin{eqnarray}
ds^2 &=&\left( 1 - \frac{r_g}{r}-\frac{1}{2}+\frac{t}{2\sqrt{t^2-r^3}} \right) dt^2\\
&-& \left( \left( 1 - \frac{r_g}{r} \right)^{-1} - \frac{1}{2}+\frac{t}{2\sqrt{t^2-r^3}} \right) dr^2,
\end{eqnarray}
in curvature coordinates. This line element develops a curvature singularity at $r=r_g$ and is strictly valid only for $r_g<r<t$, that is between the black hole horizon and the cosmological horizon. The basic behavior, building on the analysis of \cite{Babichev:2004yx} is, however, in agreement with the picture drawn in this article. Many situations beyond the radiation-dominated background were also considered (see, {\it e.g.}, \cite{Kiselev:2002dx} -- pushed further in \cite{Heydarzade:2016zof} -- which is very useful for deriving generalized metrics in a nonvacuum environment). Some interesting hints can already be found with the Vaidya metric, describing the absorption (or emission) of null dust. The very meaning of the associated horizons is, however, still debated \cite{nielsen}. The most promising approach to deal with dynamical black holes in the framework we consider is probably the one advocated in \cite{Nielsen:2005af}. Event horizons are indeed not well suited for evolving black holes, and quite a lot can be inferred from the evolution and the apparent horizon in Painlev\'e-Gullstrand coordinates (that are regular at the horizon).\\

Although the expansion (or contraction) of the Universe is not the most important aspect for the points we are making here, it would in principle make sense to generalize this study  by considering metrics found, {\it e.g.}, in \cite{Faraoni:2007es,Gao:2008jv}. In particular, when the mass depends on time in a flat FLRW background, the line element -- generalizing the McVittie results -- reads \cite{Faraoni:2008tx}

\begin{eqnarray}
ds^2&=&-\frac{\left[1-\frac{M(t)}{2a\left(t\right)r}\right]^2}{
\left[1+\frac{M(t)}{2a\left(t\right)r}\right]^2} \, dt^2
  +{a^2\left(t\right)}\left[1+\frac{M(t)}{2a\left(t\right)r}
  \right]^4
   \nonumber\\ &&\nonumber \\
  &&\times\left(dr^2+r^2d\Omega^2 \right) \;.  \label{1}
 \end{eqnarray} 
Interesting paths are also suggested in \cite{Abdalla:2013ara,Guariento:2015cxa}.\\

Two specifically interesting situations were recently considered in the case of evaporating black holes. The first one relies on the Thakurta metric \cite{Thakurta}. The Thakurta spacetime is a generalized McVittie black hole with accretion. This solution approximately corresponds (at distances such that $M \ll R \ll 1/H$) to a Newtonian point particle with growing mass,
the accretion rate being proportional to the Hubble rate. Important consequences were, {\it e.g.}, investigated in \cite{Boehm:2020jwd}, deeply changing the LIGO bounds on primordial black holes. Several concerns were raised: the energy flux required for this specific mass growth seems nonphysical
\cite{Kobakhidze:2021rsh} and neither an event horizon nor a trapping horizon seems allowed \cite{Harada:2021xze}. Answers were provided in \cite{Kobakhidze:2021rsh}, adopting the  foliation associated with the Kodama time. Other counterarguments were exhibited in \cite{Boehm:2021kzq}. Anyway, the  peculiar mass evolution, $\dot{M}=HM$, associated with this spacetime is very different from the one of our study, as even the asymptotic behaviors do not coincide \cite{Carr:1974nx,Babichev:2004yx,Babichev:2018ubo,Carr:2014pga}. 

The second one is based on the Sultana-Dyer spacetime \cite{Sultana:2005tp}. It is a Petrov type D metric describing a black hole embedded in a spatially flat FLRW universe with scale factor $a(t) \propto t^{2/3}$, generated by mapping the Schwarzschild timelike Killing field $\xi^a$ into a conformal Killing field (for $\xi^a \nabla_a\Omega\neq0$) \cite{Faraoni:2009uy}. In this framework, an exact model for evaporating primordial black holes in the cosmological spacetime was developed in \cite{Xavier:2021chn}, showing potentially important deviations with respect to the usual behavior. This approach, however, assumes a matter-dominated universe which not only is different from the one considered in this study but also is incompatible with the specific kind of accretion assumed here.\\

The situation is not fully clear. The ``correct" horizon to consider highly depends on the physical properties being investigated. Apparent horizons are usually advocated for phenomenological purposes but their foliation-dependence problem remains mostly unsolved \cite{Faraoni:2018xwo} (even though arguments in \cite{Kobakhidze:2021rsh} show that a preferred foliation of the Thakurta spacetime does exist). There is no consensus \cite{Faraoni:2018xwo} and the question of horizons in a dynamical framework is still an open one (interesting points are made in \cite{Vanzo:2011wq}).

In this work, we do {\it not} pretend to give a fully general exact and analytical solution for growing black holes in a radiation-dominated dynamical universe. This goal is still beyond available models. We simply focus on some specific features of the Schwarzschild metric in a photon background. This is obviously an oversimplification but it exhibits intriguing behaviors that deserve future investigations. As one can guess from the generalized McVittie metric (where $M(t)$ is the Hawking-Hayward quasilocal mass), it is probable that, at late times, the rate of growth of the black hole mass becomes comparable to the Hubble rate and the black hole becomes comoving \cite{Faraoni:2007es}. The situation in which such a result is derived is, however, different from the one studied in this article and no firm conclusion can be reached.

\section{Conclusion}

The behavior of black holes in a thermal bath is counterintuitive when first encountered. This is simply grounded in the very special mass/radius relation of black holes. For standard matter, the area scales as $dA\propto M^{-1/3}dM$ whereas for black holes, $dA\propto MdM$. This is the keypoint. For usual matter, the area variation induced by an incoming mass $dM$ decreases with the mass $M$ of the star (or whatever) whereas for black holes, it {\it increases} with $M$: the larger the black hole, the larger the area variation induced by the absorption of a quantum of given energy. This is the straightforward cause of the supercritical mass behavior. It is, however, mandatory to underline that all those ``strange" features, when taken literally, might very well be artifacts of the Schwarzschild metric used beyond its domain of validity more than real physical effects. More work is needed and the aim of this article is mostly to invite the interested readers to investigate in more detail the curious behaviors presented here.\\

Within those (strong) limitations, we have established that for very simple reasons, the evolution of a black hole in a thermal bath is catastrophic. If the bath is at constant temperature, the black hole mass inevitably diverges at finite time (ignoring, of course, cosmological horizon issues). If the temperature of the bath increases -- {\it e.g.} in a contracting space -- the phenomenon is (obviously) even faster and the black hole singularity is reached before the background singularity. If the temperature decreases -- {\it e.g.} in an expanding space -- the mass can either diverge or vanish, in finite time once again. For tuned initial conditions a (questionable) self-similar solution is possible. In the vanishing case, the black hole spends most of its life on a long ``plateau". 

The popular expression ``black hole bomb" \cite{Press:1972zz} is strengthened and acquires a wider meaning. 
As a possible nongravitational development of this work, it would be interesting to investigate how the studied phenomenon can be viewed as a phase transition. Analogies with an appropriate Ising model should be fruitful.

\section{Acknowledgments}

We thank Vivien Lecomte, Florent Michel, Antoine Rignon-Bret, and L\'eonard Ferdinand for very interesting remarks and helpful discussions.

\bibliography{refs.bib}

 \end{document}